\documentclass[usenatbib]{mn2e}
\usepackage{graphicx}
\usepackage{natbib}
\usepackage{amsmath}
\usepackage{dsfont}
\def\aj{{AJ}}                   
\def\apj{{ApJ}}                 
\def\apjl{{ApJ}}                
\def\apjs{{ApJS}}               
\def\ao{{Appl.~Opt.}}           
\def\aap{{A\&A}}                
\def\mnras{{MNRAS}}             
\def\prd{{Phys.~Rev.~D}}        
\def\nat{{Nature}}              

\begin{document}

\title[Non-linear contribution to the ISW effect]{Towards accurate modelling of
the ISW effect, the non-linear contribution} 
\author[Cai et al.]{Yan-Chuan Cai, Shaun Cole, Adrian Jenkins, and Carlos Frenk \\
Institute for
Computational Cosmology, Durham University, South Road, Durham, UK}
\maketitle
\begin{abstract}
In a universe with a cosmological constant, the large-scale
gravitational potential varies in time and this is, in principle,
observable. Using an N-body simulation of a $\Lambda$CDM universe, we
show that linear theory is not sufficiently accurate to predict the
power spectrum of the time derivative, $\dot{\Phi}$, needed to compute
the imprint of large-scale structure on the cosmic microwave
background (CMB). The linear part of the $\dot{\Phi}$ power spectrum
(the integrated Sachs-Wolfe effect or ISW) drops quickly as the
relative importance of $\Omega_{\Lambda}$ diminishes at high redshift,
while the non-linear part (the Rees-Sciama effect or RS) evolves more
slowly with redshift. Therefore, the deviation of the total power
spectrum from linear theory occurs at larger scales at higher
redshifts. The deviation occurs at $k\sim 0.1 $ $h$~Mpc$^{-1}$ at
$z=0$. The cross-correlation power spectrum of the density $\delta$
with $\dot{\Phi}$ behaves differently to the power spectrum of
$\dot{\Phi}$. Firstly, the deviation from linear theory occurs at
smaller scales ($k\sim 1 $ $h$~Mpc$^{-1}$ at $z=0$). Secondly, the
correlation becomes negative when the non-linear effect dominates. For
the cross-correlation power spectrum of galaxy samples with the CMB,
the non-linear effect becomes significant at $l\sim 500$ and rapidly
makes the cross power spectrum negative. For high redshift samples,
the cross-correlation is expected to be suppressed by $5-10\%$ on
arcminute scales. The RS effect makes a negligible contribution to the
large-scale ISW cross-correlation measurement. However, on arc-minute
scales it will contaminate the expected cross-correlation signal
induced by the Sunyaev-Zel'dovich effect.
\end{abstract}
\section{Introduction}

The most intriguing topic in contemporary cosmology is the nature of the dark
energy which appears to dominate the energy density of the Universe at late
times.  Strong evidence for the existence of dark energy comes from both the
combined analysis of the cosmic microwave background radiation (CMB) and
the galaxy large-scale structure (LSS) 
\citep[e.g.][]{Efstathiou02,Spergel03}, and from high redshift type~Ia supernovae
\citep[e.g.][]{Riess98,Perlmutter99}.  Both of
these techniques infer the presence of dark energy from geometrical
measures. A complementary probe of dark energy is provided by
techniques that measure the dynamical effect of dark energy through
its influence on the rate of growth of structure.  Large deep galaxy
redshift surveys (like the EUCLID, the ESA Mission to Map the Dark
Universe, and the JDEM, the Joint Dark Energy Mission) are being
planned that will exploit the redshift space anisotropy of galaxy
clustering, caused by coherent flows into overdense regions and
outflows from underdense regions, to measure directly the growth rate
as a function of redshift.  

The Integrated Sachs-Wolfe (ISW) effect \citep{Sachs67}, in which the
decay of the large-scale potential fluctuations induces CMB
temperature perturbations, provides another measure of the dynamical
effect of dark energy.  In principle, the ISW effect could be detected
directly in the CMB power spectrum at very low multiples.  In the
$\Lambda$CDM cosmology, it would boost the plateau in the power
spectrum at $l \sim 10$. However, as the increase of the power is not
large in comparison to the cosmic variance, it cannot be unambiguously
detected even in the WMAP data \citep{Hinshaw08}. A more sensitive
technique is to search for the ISW signal in the cross-correlation of
the LSS with the CMB. As the expected signal is weak and occurs on
large scales, a very large galaxy survey is needed to trace the
LSS. Currently individual detections based on surveys such as APM,
2MASS, NVSS and SDSS
\citep[e.g.][]{Fosalba03,Afshordi04,Fosalba04, Padmanabhan05, Cabre06,McEwen07,
Rassat07, Raccanelli08} are not of very high statistical significance \citep[but
see][]{Granett08}. There are also analyses of the ISW cross-correlation, which 
combine multiple galaxy survey samples and achieve $\sim 4\sigma$ detection of 
the ISW effect (Ho et al. 2008; Giannantonio et al. 2008). 
These measurements may be the best that can be obtained before the next 
generation of surveys (BOSS\footnote{http://www.sdss3.org/cosmology.php}, 
Pan-STARRS1\footnote{http://www.ps1sc.org/}) come to fruition and make redshift tomography 
possible. If such surveys are to place robust, meaningful
constraints on the properties of the dark energy it is important to take full
account of other processes beyond the (linear) ISW effect that may contribute to
the cross-correlation signal. Here, we focus on deviations caused by non-linear
gravitational evolution, the Rees-Sciama effect \citep{Rees68}.

Other processes are known to contribute to the cross-correlation
signal. First, the thermal Sunyaev-Zel'dovich (SZ) effect
\citep{Sunyaev72} caused by hot ionized gas in galaxy clusters induces
an anti-cross-correlation signal which can cancel the ISW effect on
small scales. Its statistical contribution can be modelled and
subtracted given the value of $\sigma_8$ (the $rms$ linear mass
fluctuations within a sphere of 8 $h^{-1}$~Mpc) which determines the
abundance of galaxy clusters
\citep[e.g.][]{White93,Fan01,Mei04}. Also, since the thermal SZ effect is frequency
dependent, it can be subtracted in frequency space given sufficient spectral
coverage.  Second, the redshift dependence of galaxy bias, if not properly taken
into account, can introduce systematic effects in the determination of dark
energy parameters. Other effects such as lensing magnification and the Doppler
redshift effect can also boost the cross-correlation signal, but are only
important at high redshift \citep{Loverde07,Giannantonio07}. These effects are
well documented and can be calibrated and removed.

In this paper we will solely explore the contribution of the non-linear terms,
or the Rees-Sciama (RS) effect, on the cross-correlation signal.  The RS effect
arises from the non-linear evolution of the potential \citep{Rees68}. It is 
believed to be much smaller than the CMB signal at all
scales \citep{Seljak96, Puchades06}. Indeed, compared with the CMB power
spectrum, the RS effect is orders of magnitude lower. Also, compared with the
complete integrated ISW power spectrum, the RS effect has been shown, using the
halo model approach \citep[e.g.][]{Cooray02a, Cooray02b}, to be unimportant at
$l<100$. However, the RS effect has not been taken into account in
cross-correlation analyses and it is important to assess its importance ahead of
the completion of the next generation of large deep galaxy surveys.

We use a large N-body simulation to investigate the effect of the non-linear
contribution on the interpretation of the ISW cross-correlation signal.  We use
the $488^3$-particle L-BASICC simulation described by
\cite{Angulo08} which, with a box size of 1340 $h^{-1}$~Mpc, is ideal for this
purpose because not only does it enable us to extrapolate our analysis to
non-linear scales at different redshifts, but it includes the very large scale
power necessary to check the agreement with linear theory. The cosmology adopted
in the L-BASICC simulation is $\Lambda$CDM, with $\Omega_\Lambda=0.75$,
$\Omega_{\rm m}=0.25$, $\Omega_{\rm b}=0.024$, $\sigma_8=0.9$ and
$H_0=73$~km~s$^{-1}$~Mpc$^{-1}$.

The paper is organised as follows. In \S2, we compute the power spectrum of the
ISW plus RS effects from our simulation and compare them with linear theory. In
\S3, we analyse these two effects in terms of the cross-correlation of the LSS with
the CMB. Finally, in \S4, we discuss our results and present our conclusions.

\section{\label{sec:level1}Time derivative of the potential}
The integrated Sachs-Wolfe effect results from the late time decay
of gravitational potential fluctuations. The net blueshift or redshift
of the CMB photons caused by the change in the potential during the
passage of the photons induces net temperature fluctuations of the
black body spectrum,
\begin{equation}\label{eq1}
\frac{\Delta T (\hat n)}{\bar{T_0}} = -\frac{2}{c^2}\int_{0}^{t_L}
\dot{\Phi}(t,\hat n) \, dt,
\end{equation}
where $\dot{\Phi}$ is the time derivative of the gravitational
potential, $t$ is the lookback time, with $t=0$ at the present and
$t=t_L$  at the last scattering surface.
The angular power spectrum of these temperature fluctuations
(see Appendix A) is given by
\begin{equation}{\label{eq2}}
\begin{aligned}
C_l=\frac{4}{c^4}\frac{2}{\pi} \int \int_0 ^{t_{L}}\int_0 ^{t_{L}} k^2 P_{\dot\Phi\dot\Phi}(k, r, r')j_l(kr)j_l(kr')dt' dt d k \\
 \approx \frac{4}{c^4}\int_0 ^{t_{L}} P_{\dot\Phi\dot\Phi}(k=\frac{l}{r}, t)/r^2\, dt ,
\end{aligned}
\end{equation}
where $r$ is the comoving distance to lookback time, $t$, $j_l$ is the
spherical Bessel function and $P_{\dot{\Phi}\dot{\Phi}}(k,r)$ is the
3-D power spectrum of $\dot{\Phi}$ fluctuations.  To derive the final
expression we have used Limber's approximation by assuming $k \approx
l/r$ \citep[][also see Appendix A]{Limber54,Kaiser92,Hu00,Verde00}.

The ISW effect consists of the temperature fluctuations described by these
equations when linear theory is used to compute $\dot\Phi$ and its fluctuation
power spectrum $P_{\dot{\Phi} \dot{\Phi}}$.  Using a simulation to determine the
non-linear contributions we can quantify the full ISW plus Rees-Sciama effect.
In Fourier space, the time derivative of the gravitational potential can be
expressed as:
\begin{equation}\label{eq3}
\dot{\Phi}(\vec{k},t)=\frac{3}{2}\left(\frac{H_0}{k}\right)^2\Omega_{\rm m}
\left[\frac{\dot{a}}{a^2}\delta(\vec{k},t)-\frac{\dot{\delta}(\vec{k},t)}{a}\right],
\end{equation}
where $a$ is the expansion factor, $H_0$ is the Hubble constant,
$\Omega_{\rm m}$ is the present mass density parameter and
$\dot{\delta}$ is the time derivative of the
density fluctuation. Combining this with the Fourier space form of
the continuity equation, $\dot\delta(k,t)+i\vec k\cdot\vec{p}(\vec{k},t)=0 $
gives:
\begin{equation}\label{eq4}
\dot{\Phi}(\vec{k},t)=\frac{3}{2}\left(\frac{H_0}{k}\right)^2\Omega_{\rm m}
\left[\frac{\dot{a}}{a^2}\delta(\vec{k},t)+\frac{i\vec{k}\cdot\vec{p}(\vec{k},t)}{a}\right],
\end{equation}
where $\vec{p}(\vec{k},t)=[1+\delta(\vec{k},t)]v(\vec{k},t)$ is the momentum
density field in Fourier space divided by the mean mass density.  This enables
us to estimate the Fourier transform of the $\dot{\Phi}$ field of the simulation
from the Fourier transforms of the density and momentum fields. Using
equation~(\ref{eq3}), the resulting power spectrum,
$P_{\dot{\Phi}\dot{\Phi}}(k,t)= (2\pi)^{-3} \langle \dot{\Phi}(k,t)
\dot{\Phi}^*(k,t) \rangle$, can be written as
\begin{equation}\label{eq6}
\begin{aligned}
P_{\dot{\Phi}\dot{\Phi}}(k,t)=\frac{9}{4}\left(\frac{H_0}{k}\right)^4\Omega_{\rm m}^2 \times
 \ \ \ \ \ \ \ \ \ \ \ \ \\
\left[\left(\frac{\dot{a}}{a^2}\right)^2 P_{\delta\delta}(k,t)-2\frac{\dot{a}}{a^3}
P_{\delta \dot{\delta}}(k,t)+\frac{1}{a^2}P_{\dot{\delta}
\dot{\delta}}(k,t)\right] .
\end{aligned}
\end{equation}
\begin{figure}
\resizebox{\hsize}{!}{
\includegraphics{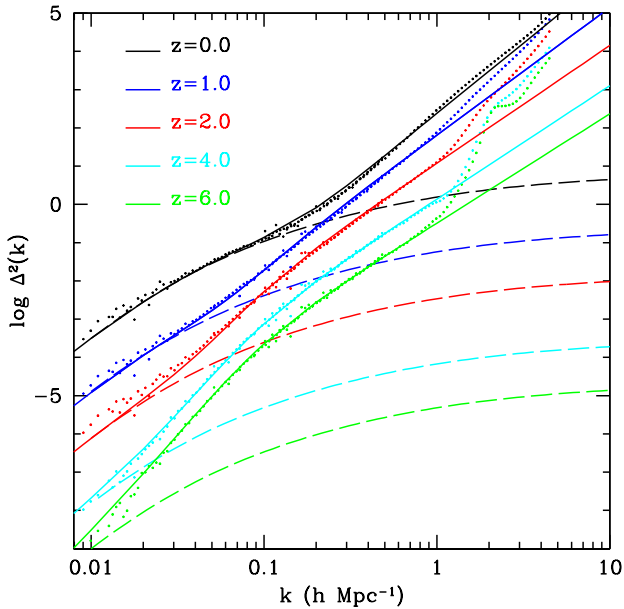}}
\caption{\label{fig1}Scaled $\dot{\Phi}$ power spectrum,
$\mathcal{P}_{\dot{\Phi}\dot{\Phi}}(k)$, at different redshifts,
$\Delta^2(k)\equiv k^3\mathcal{P}_{\dot{\Phi}\dot{\Phi}}(k)/2\pi^2$. The dotted
lines are the measurements from the L-BASICC simulation. The
solid lines are our model, while the dashed lines are the linear theory. The
deviation of the simulation results from linear theory happens at larger scales
as redshift increases. We also find that the deviation from linear theory for
the $\dot{\Phi}$ field occurs at larger scales than that of the density field at
all redshifts.}
\end{figure}
\begin{figure}
\resizebox{\hsize}{!}{
\includegraphics{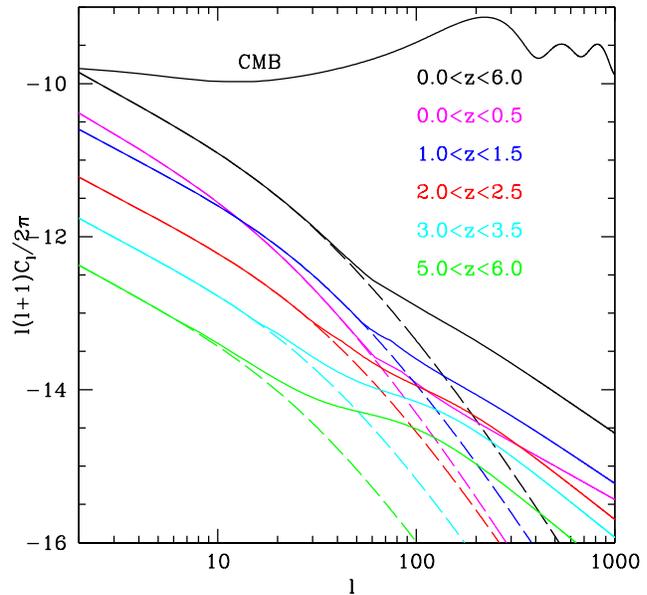}}
\caption{\label{fig2}ISW angular power spectrum coming from different
redshift intervals, where $l(l+1)C_l/2\pi = (\Delta T/T)^2$.  The
solid lines are given by Eq. (\ref{eq2}) and Eq. (\ref{eq6}),
evaluated using our model of the measurements from the L-BASICC
simulation. The dashed lines are linear theory. The power spectrum at
$0<z<6$ shows that the deviation of the simulation results from linear
theory starts at $l<100$. The deviation starts at smaller $l$ as
redshift increases.}
\end{figure}

In linear theory,
$P_{\dot{\delta}\dot{\delta}}(k,t)=k^2P_{vv}(k,t)=\dot{D}(t)^2P_{\delta\delta}^{\rm
{lin}}(k)$ and $P_{\delta\dot{\delta}}(k,t)=kP_{\delta
v}(k,t)=D(t)\dot{D}(t)P_{\delta\delta}^{\rm {lin}}(k)$, where
$P_{\delta\delta}^{\rm{lin}}(k)$ is the linear density power spectrum at the
present time and $D(t)$ is the growth factor normalised to be unity at
present. Therefore, the power spectrum of the linear ISW effect is
\begin{equation}
P_{\dot{\Phi}\dot{\Phi}}^{\rm{lin}}(k,t)=\frac{9}{4}\left(\frac{H_0}{k}\right)^4\Omega_{\rm m}^2
\left[\frac{H(t)D(t)(1-\beta)}{a}\right]^2P_{\delta\delta}^{\rm{lin}}(k),
\end{equation}
where $\beta=\frac{{\rm d}\ln D}{{\rm d}\ln a}\simeq\Omega_{\rm m}^{0.6}(t)$.
For easy comparison at different redshifts in the simulation, we defined a 
scaled $\dot{\Phi}$ power spectrum, 
$\mathcal{P}_{\dot{\Phi}\dot{\Phi}}=
P_{\dot{\Phi}\dot{\Phi}}/[\frac{9}{4}\left(\frac{H_0}{k}\right)^4\Omega_{\rm
    m}^2]$
which from Eq. (5) is simply
\begin{equation}
\mathcal{P}_{\dot{\Phi}\dot{\Phi}}(k,z)\equiv
P_{\delta\delta}(k,z)-2\frac{P_{\delta
    \dot{\delta}}(k,z)}{H(z)}+\frac{P_{\dot{\delta}\dot{\delta}}(k,z)}{H^2(z)} .
\end{equation}
Our measurements of the $\mathcal{P}_{\dot{\Phi}\dot{\Phi}}(k,z)$
power spectrum are shown in Fig.~\ref{fig1}. The results from linear theory
are also plotted. We find the total scaled $\dot{\Phi}$ power spectrum can be
well fitted by a broken power law plus the linear scaled $\dot{\Phi}$ power
spectrum $\mathcal{P}_{\dot{\Phi}\dot{\Phi}}
=\mathcal P_{\dot{\Phi}\dot{\Phi}}^{\rm{nonlin}}+
\mathcal P_{\dot{\Phi}\dot{\Phi}}^{\rm{lin}}$, where
\begin{equation}
\mathcal P_{\dot{\Phi}\dot{\Phi}}^{\rm{nonlin}}(k,z)=\frac{A}{(10k)^{-4/0.75}+(10k)^{-4/B}}.
\end{equation}
Here A and B are two free parameters that we use to fit the  model to
the simulation results at each redshift up to $z=6$. To interpolate
the model to intermediate redshifts we linearly interpolate the values
of A and B from the nearest two simulation outputs. Our model is
compared to the simulation results in Fig. \ref{fig1}.
\begin{figure*}
\resizebox{\hsize}{!}{
\includegraphics{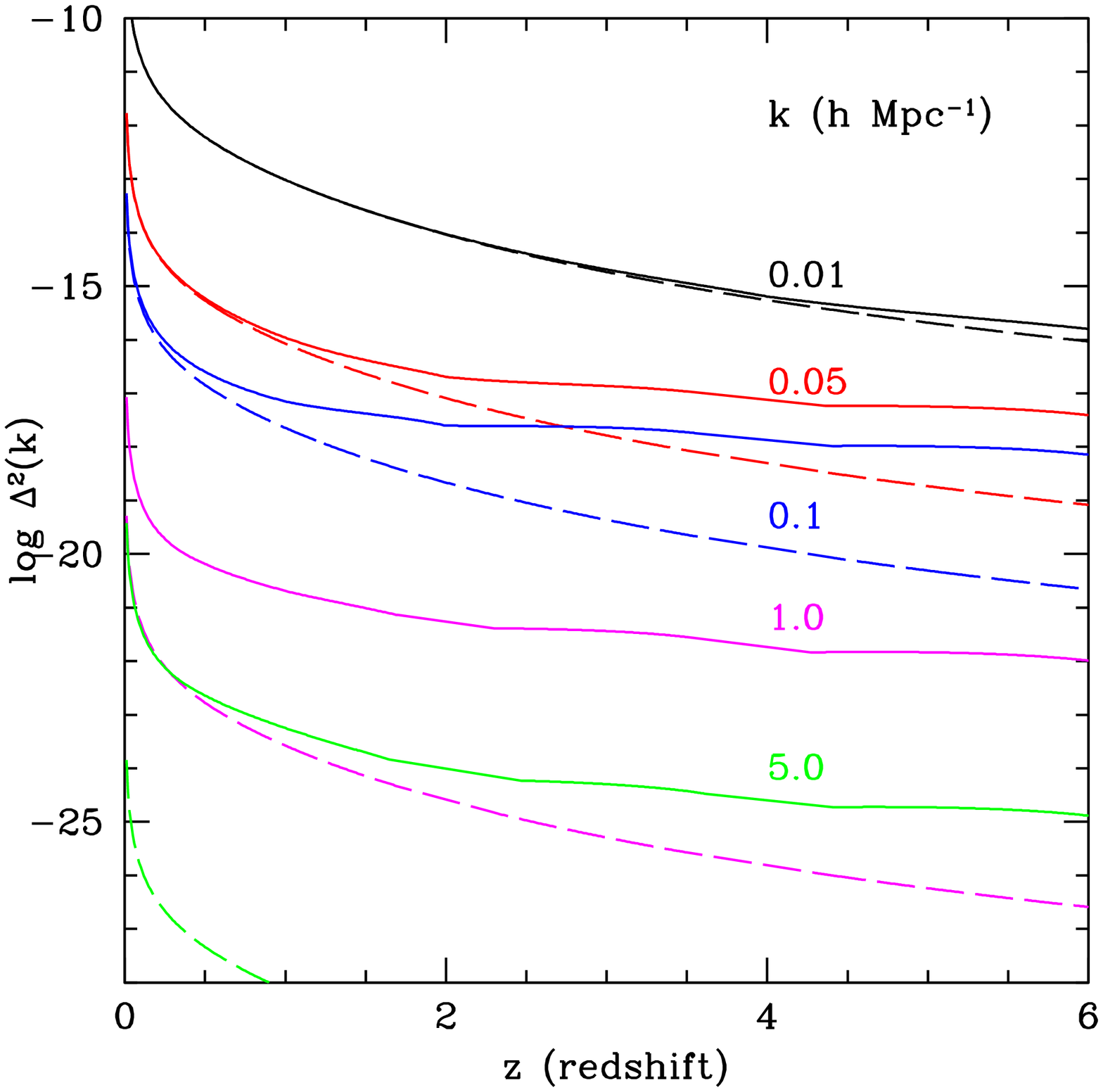}
\includegraphics{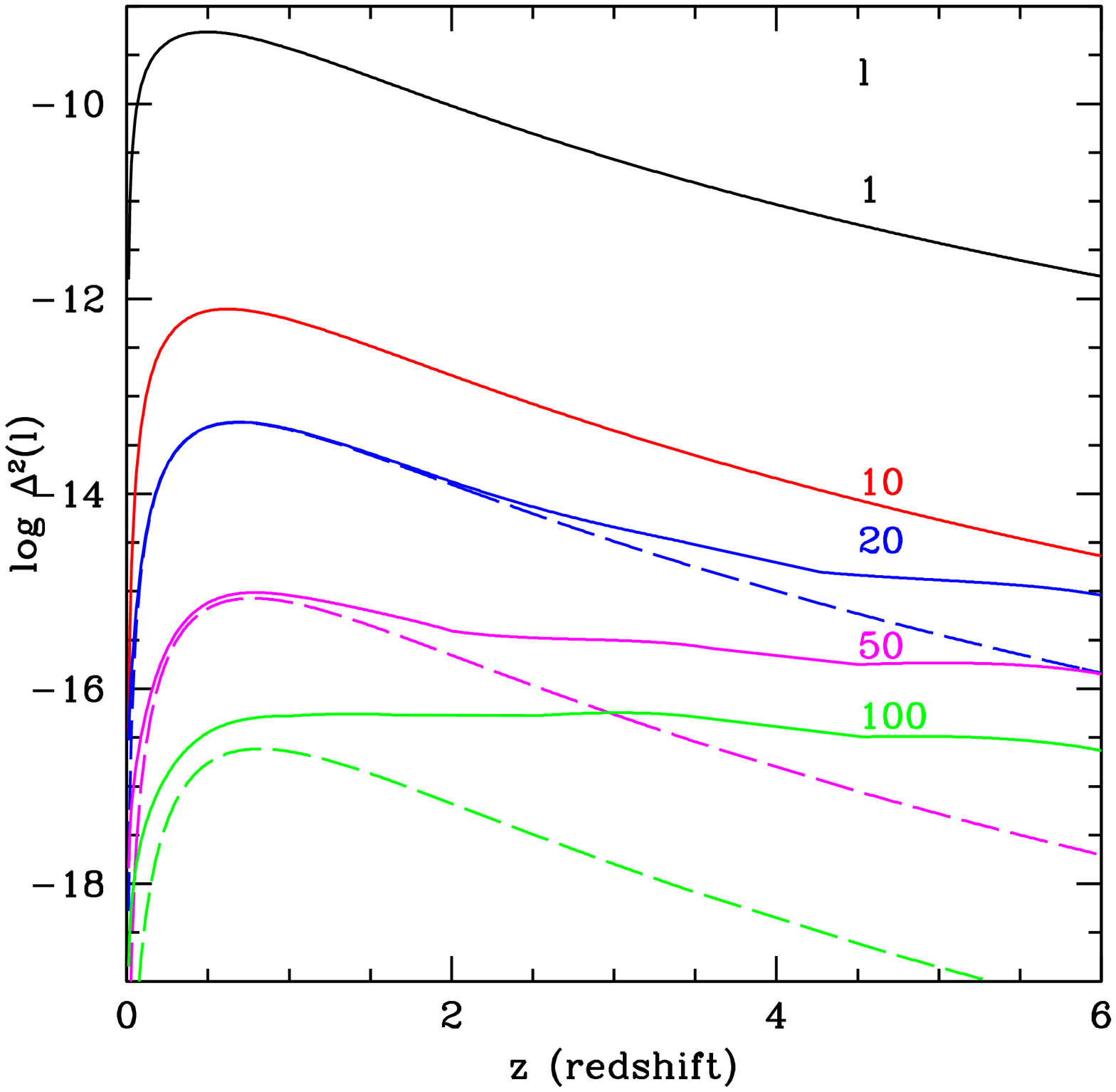}}
\caption{\label{fig3} Evolution of the $\dot{\Phi}$ power spectrum 
$\Delta^2(k)\equiv k^3 \mathcal{P}_{\dot{\Phi}\dot{\Phi}}(k)/2\pi^2$ for 
specific spatial (left) and angular (right) modes. The solid lines show our 
model of the simulation results. The dashed lines show the results of linear 
theory. The power $\Delta^2(k)$ decreases monotonically as a function 
of $z$ at all scales in linear theory. However, the total power seems to be 
independent of $z$ at high redshift. The deviation from linear theory 
increases with $z$ and $k$. The angular power spectrum $\Delta^2(l)$ shows 
no deviation from linear theory up to $l=10$. For $l>50$, deviations 
appear at all redshifts.}
\end{figure*}

We see in Fig.~\ref{fig1} that the linear theory reproduces the ISW+RS
$\mathcal{P}_{\dot{\Phi}\dot{\Phi}}$ at $z=0$ only at $k<0.1$
$h$~Mpc$^{-1}$.  It fails at progressively larger scales as the
redshift increases. By $z=2$, linear theory agrees with the simulation
results only at $k<0.02$ $h$~Mpc$^{-1}$. The reason for this
surprising behaviour is that the linear part of the
$\mathcal{P}_{\dot{\Phi}\dot{\Phi}}$ drops quickly to zero as the
relative importance of $\Omega_{\Lambda}$ diminishes at high redshift,
while the non-linear part evolves more slowly with
redshift. Therefore, the deviation of the total power spectrum from
linear theory happens at larger scales at higher redshifts. We find
that the momentum power spectrum, $P_{\dot{\delta}\dot{\delta}}$, and
the correlation power spectrum of the density and momentum,
$P_{\delta\dot{\delta}}$, behave similarly to the
$P_{\dot{\Phi}\dot{\Phi}}$ power spectrum, namely, their deviation
from linear theory occurs at larger scales at higher redshift. This is
in contrast with the power spectrum of the density field which deviates 
from linear theory on progressively larger scales at lower and lower 
redshift. In
another words, at the same redshift, the deviation from linear theory
occurs at smaller scales for the density field than for the other
fields.

The sharp increase of $\mathcal{P}_{\dot{\Phi}\dot{\Phi}}$ measured from the
simulation at small scales ($k>1$ $h$~Mpc$^{-1}$) is due to discreteness in the
$448^3$ particle L-BASICC simulation.  We used the much higher resolution
$2160^3$ particle Millennium simulation \citep{Springel05} to verify that our
model remains accurate at smaller scales and is robust to shot noise
corrections.

We can now compute the induced angular power spectrum of CMB temperature
fluctuations by performing the integral in equation~(2) over the redshift range
$0<z<6$ using our model for the 3-D power spectrum, $P_{\dot \Phi \dot
\Phi}(k,z)$. The overall result is shown in Fig.~\ref{fig2} along with the
contributions coming from different redshift intervals.  For the overall angular
power spectrum the deviation of the model from the linear theory happens at $l
\sim $100. This result confirms the prediction of
\citet{Cooray02b} based on the halo model.
However, we also see that the failure of linear theory, as judged by our
simulation results, occurs at smaller and smaller $l$ as redshift increases.
For example, above $z=5$, the deviation occurs at $l<20$ and, for larger values
of $l$ than this, linear theory becomes extremely inaccurate.

In order to evaluate how the breakdown of linear theory depends on redshift, we
plot the evolution of the $\dot{\Phi}$ power at a given scale as a function of
redshift in Fig.~\ref{fig3}.  Generally, the deviations of linear theory from
the simulation results decrease with scale and increase with redshift.  At
$k=0.01$ $h$ Mpc$^{-1}$, deviations start to be seen at $z \sim 3$ and, at
$k=0.1$ $h$ Mpc$^{-1}$, linear theory has become inaccurate at all redshifts. In
the right-hand panel, which shows results in $l$ space, we find no deviations up
to $l\sim 10$, but for $l>50$, linear theory has clearly broken down at all
redshifts.  Interestingly, at high redshift, the $\dot{\Phi}$ power in the
simulation appears to be independent of $z$ while, in linear theory, this
quantity drops monotonically with $z$.

\section{the LSS-CMB cross-correlation}
\begin{figure*}
\resizebox{\hsize}{!}{
\includegraphics{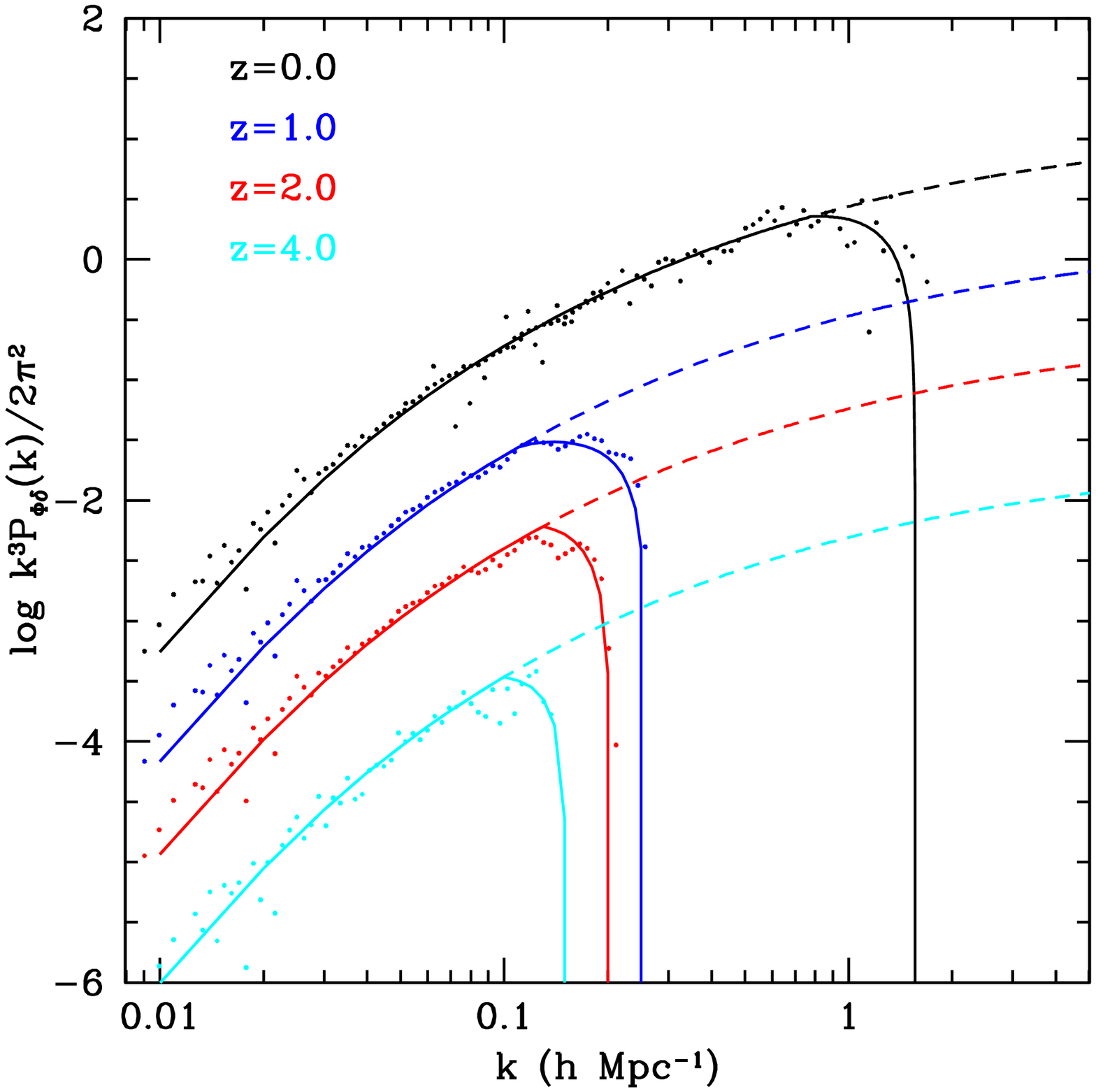}
\includegraphics{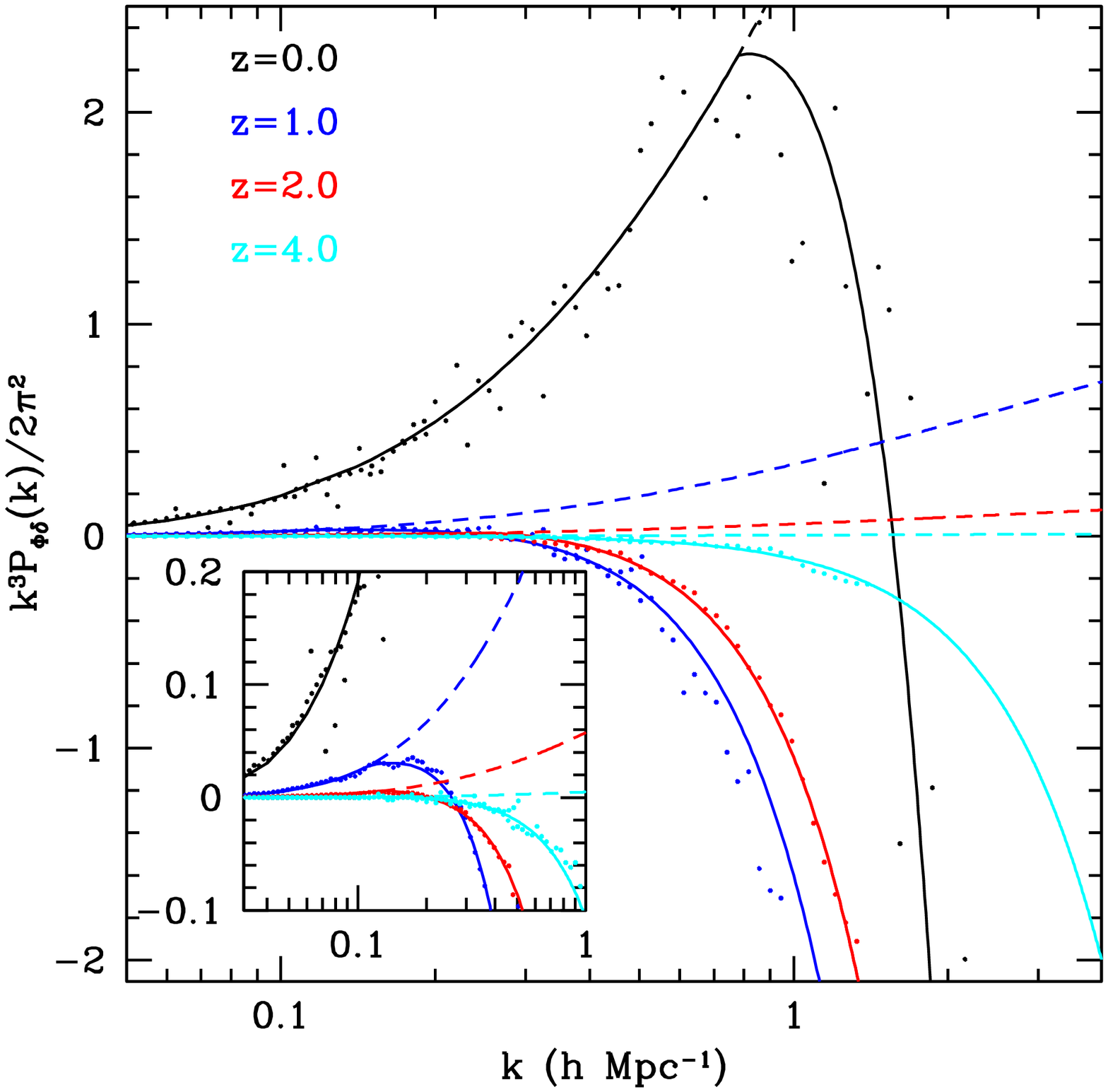}}
\caption{\label{fig3.1}$Left$: The cross power spectrum of $\dot{\Phi}$ with 
$\delta$ at different redshifts. The dotted lines represent the
measurements from the L-BASICC and the dashed lines linear theory. The
solid lines are our model fit to the sum of the linear theory and the
non-linear contribution in the L-BASICC simulation.  The non-linear
effect begins to appear at $k\sim 1$ $h$ Mpc$^{-1}$ at $z=0$, and at
larger scales at higher redshifts. It rapidly suppresses the cross
power spectrum and causes it to become negative on small
scales. $Right$: the same as the left panel but on a linear scale.}
\end{figure*}
To illustrate the contribution of the Rees-Sciama effect to the 
cross-correlation of the density field $\delta$ with $\dot{\Phi}$, we can 
compute the 3-D cross-correlation power spectrum $P_{\dot{\Phi}\delta}(k,z)=
<\dot{\Phi}(\vec k, z)\delta^*(\vec k, z)>$ from our simulations:
\begin{equation}
\mathcal P_{\dot{\Phi}\delta}(k,z)=P_{\delta\delta}(k,z)-\frac{P_{\dot{\delta}\delta}(k,z)}{H(z)}
\end{equation}
In linear theory, $\mathcal
P_{\dot{\Phi}\delta}(k,z)=D^2(1-\beta)P_{\delta\delta}^{\rm{lin}}(k)$,
where $P_{\delta\delta}^{\rm{lin}}(k)$ is the linear density power
spectrum at the present time.  Results from our simulations are shown
in Fig.~\ref{fig3.1}. Comparing with linear theory, we find that the
non-linear contribution appears at somewhat smaller scales than that
of the autocorrelation power spectrum of $\dot{\Phi}$. At $z=0$, the
deviation occurs at $k\sim 1$ $h$ Mpc$^{-1}$. However, it then
dominates rapidly, making the cross-correlation power spectrum
negative. This indicates that once the non-linear effect dominates,
the potential of overdense regions evolves faster than the expansion
of the universe, becoming deeper and thus imparting a net redshift to
CMB photons passing through them. The sense of the effect from 
underdense regions is reversed, but the effect generated by the overdense 
regions is dominant and induces a negative cross-correlation between the 
CMB and the LSS.  \footnote{An alternative way of seeing the sign of the 
Rees-Sciama effect is to consider the cross power spectrum $P_{\dot\Phi\delta}$
which from Eq.~(9) can be seen to be proportional to
$d( P_{\delta\delta}/a^2)/dt$. The ISW effect is the result of the
linear density power spectrum growing less rapidly than $a^2$ once
curvature or $\Lambda$ become dominant. In contrast the Rees-Sciama
effect results from the non-linear growth of the density power
spectrum, which for the times and scales of interest grows faster than
linear theory \citep[e.g.][]{Smith03}. Consequently the
contribution to the cross power spectrum from
the Rees-Sciama effect has the opposite sign to that of the
ISW effect.}

To quantify the effect on the
large-scale ISW cross-correlation measurements, we model the 3-D cross
power spectrum from the simulations at each redshift output. We use
linear theory to model the linear regime but once the cross power
spectrum starts to deviate from linear theory, we fit it with a
function of the form $k^3P_{\dot{\Phi}\delta}(k)=A_1+A_2k+A_3k^2$,
where $A_1, A_2$ and $ A_3$ are free parameters.  Interpolating to
intermediate redshifts, we are then able to calculate the projected 2-D
power spectrum.

The cross-correlation between LSS and CMB maps has been shown to be a powerful
tool for verifying the existence of dark energy and constraining its
properties. Current measurements of the cross-correlation have low statistical
significance because the volumes probed by LSS surveys are relatively small, but
this situation will improve greatly with upcoming surveys. For example,
Pan-STARRS1 will survey three quarters of the sky, obtaining photometry for
galaxies up to $\sim 24.6$ mag in the $g$-band. The mean galaxy redshift in this
``$3\pi$ survey'' will be $\bar z\sim 0.5$. Pan-STARRS1 will also carry out a
deeper but smaller ``MDS'' survey covering 84 sq deg of the sky to $\sim 27.3$
mag in $g$ for which $\bar z\sim 0.8$ \citep{2008Cai}. Cross-correlating such
photometric redshift galaxy samples with a CMB map (from WMAP or Planck) will
make it possible for the first time to perform ISW tomography. Galaxy samples
would be divided into different redshift slices and each one cross-correlated
with the CMB map. Values of the dark energy equation of state parameter, $w$,
could then be measured using the results from the different redshift slices,
effectively constraining the evolution of $w$.

To illustrate how ISW tomography may work, we follow \citet{Baugh93} and model
the redshift distribution of galaxies tracing the LSS as
\begin{equation}
N(z) \propto
\begin{cases}  (z-z_c)^2 \exp\left[-(\frac{z-z_c}{z_0})^{3/2}\right]
&\text{if $z\le z_c$} \\
 0 &\text{if $z> z_c$} ,
\end{cases}
\end{equation}
but then choose the parameters $z_0$ and $z_c$ to emulate plausible photometric
redshift slices.  (The same functional form was also taken by
\cite{Cabre06} to model the SDSS LRG sample.) We assume $z_0=0.2$ so that the
width of $N(z)$ is much greater than the expected photometric redshift errors. We shift
the function into different redshift intervals by using $z_c=0, 1, 2$ and $3$.
The median redshift of these samples is $\bar z \approx 1.4z_0+z_c=0.28,
1.28, 2.28$ and $3.28$ respectively.
The cross-correlation power spectrum (derived in an analogous way to the
auto-correlation function detailed in the appendix) is given as:
\begin{equation}
C_l^{\dot{\Phi}-g}\approx \frac{2}{c^2}\int_0^{z_L}
P_{\dot{\Phi}\delta}(k=\frac{l}{r}, z)b(z)N(z)H(z)/r^2 dz, 
\end{equation}
where $P_{\dot{\Phi} \delta}(k,z)$ is the cross power spectrum of the potential
field and the galaxy density field, $b(z)$ is the galaxy bias parameter at
redshift $z$, and $N(z)$ is the normalised galaxy selection function, where
$\int N(z)dz=1$. We adopt the small angle approximation in which $k=l/r(z)$, 
where $r(z)$ is the comoving distance.  For simplicity, in this illustration, 
we assume the galaxy bias parameter to be unity.  In angular space, the 
cross-correlation becomes:
\begin{equation}
w^{\dot{\Phi}-g}(\theta)=\sum_l \frac{2l+1}{4\pi}P_l(\cos
\theta)C^{\dot{\Phi}-g}_l, 
\end{equation}
where $P_l$ are Legendre polynomials. In actual measurements of CMB
fluctuations, the monopole and dipole are subtracted. Therefore, we set the
power at $l=0$ and $l=1$ to zero before converting the signal into real
space. To ensure that the results at smaller angles ($\theta<1^{\circ}$) converge
accurately, we sum the power up to $l=10\, 000$.
\begin{figure*}
\resizebox{\hsize}{!}{
\includegraphics{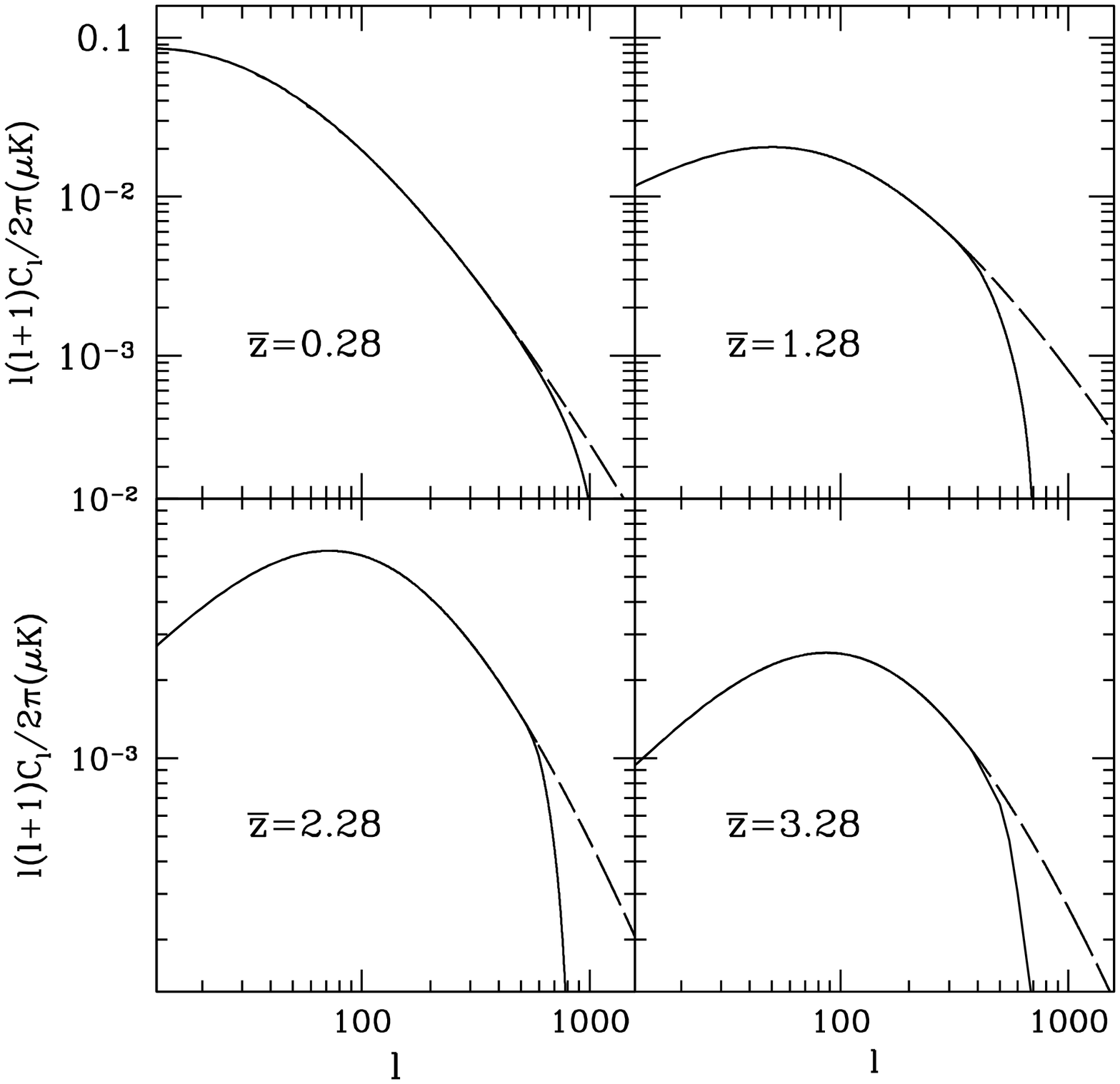}
\includegraphics{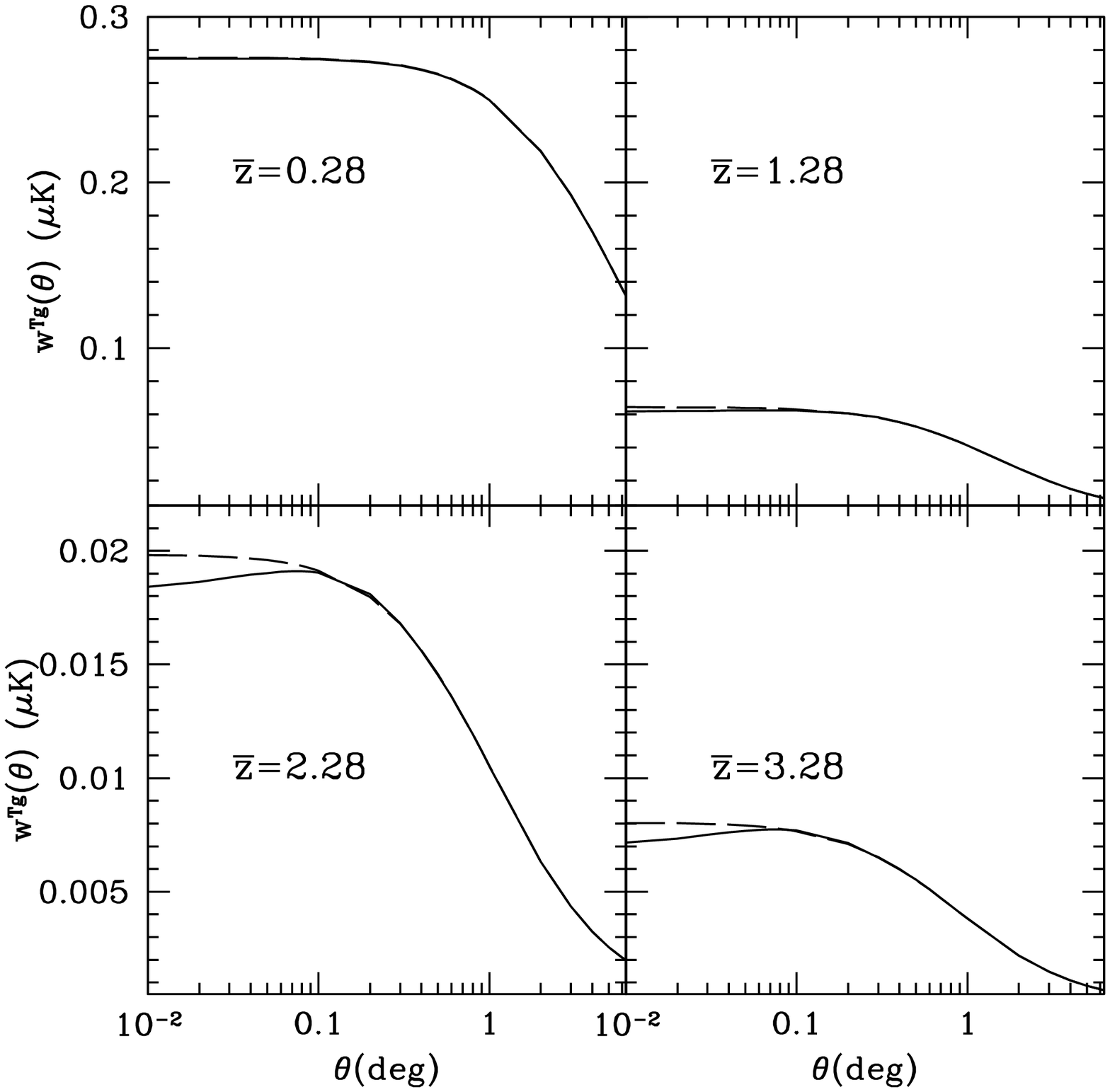}}
\caption{\label{fig4}$Left$: The cross-correlation power spectrum of galaxy samples
at different $\bar z$ with the CMB. The dashed lines are given by
linear theory. The solid lines are the sum of the linear theory and
the non-linear contribution, which is given by our model fitted to the
L-BASICC simulation.  The non-linear effect begins to appear at $l\sim
500$. It rapidly makes the cross power spectrum become
negative. $Right$: The cross-correlation of galaxy samples at
different $\bar z$ with the CMB in angular space. The non-linear
effect suppresses the ISW effect at sub-degree scales. It is
negligible at low redshifts and large scales.}
\end{figure*}

The cross-correlation results are shown in Fig.~\ref{fig4}. The
contribution from the non-linear RS effect can be seen to become
increasingly important as the redshift of the sample increases. The
cross-correlation power spectrum decreases and deviates from linear
theory rapidly at $l\sim 500$ due to the non-linear effect. It turns
negative at $l\sim 1000$. This result is consistent with that obtained
by \citet{Nishizawa08} who used a similar method to illustrate the
impact of the ISW and RS effects on the CMB-weak lensing
cross-correlation.  In angular coordinates, shown on the right-hand
panel, the RS effect is negligible at $\theta>1^{\circ}$  at all
redshifts. For the high redshift samples, it becomes important at
sub-degree scales for the high redshift samples where it suppresses
the cross-correlation power spectrum by about $5-10\%$ at acrminute
scales.

The statistical significance of current measurements of the CMB-LSS
cross-correlation is not yet high enough to detect the effects we are
discussing. Most current measurements can only determine the 
cross-correlation at degree scales or above
\citep[e.g.][]{Cabre06}. Future CMB or SZ surveys with high resolution
might be able to resolve this contribution.

\section{Conclusions}

We have used an N-body simulation to calculate the non-linear
(Rees-Sciama) contribution to the Integrated Sachs Wolfe effect. The
comparison of the 3-D and 2-D power spectra measured from the
simulation with those given by linear theory reveals a strong
nonlinear contribution whose physical scale increase
with redshift. We investigated the strength of this effect on the
cross-correlation of the CMB with galaxy samples in terms of angular
power spectra and in angular coordinates at different redshifts. We
find that there is a non-linear contribution to the cross-correlation
signal at sub-degree scales. The non-linear effect alters not only the
amplitude, but also the shape of the cross-correlation power
spectrum. With current galaxy samples which cover relatively small
volumes, it is not yet possible to disentangle the contribution of the
RS effect from that of the ISW effect within the noise. However, in
future surveys like Pan-STARRS and LSST, for which the number of
galaxies and the sky coverage will increase dramatically, the error
bars on the cross-correlation will be much smaller. In this case, the
importance of the Rees-Sciama effect may become significant for high
redshift samples.  The effect of the non-linear cross-correlation at
scales of arcminutes would contaminate the SZ signal in the CMB and
this could confuse its interpretation
\citep[e.g.][]{Fosalba03,Diego03,Myers04, Lieu06, Cao06,Bielby07}.

Our analysis is based on a simulation that assumes a $\Lambda$CDM cosmology. The
non-linear contribution depends on the values of the cosmological parameters. In
a flat universe with a cosmological constant, the RS effect will become
increasingly dominant relative to the ISW effect as the value of
$\Omega_{\Lambda}$ decreases. In the most extreme case, if
$\Omega_{\Lambda}=0.0$, the ISW effect will vanish, leaving only the RS effect
\citep{Seljak96}.  The analysis of this paper could be generalized
either using re-normalised 
perturbation theory \citep[e.g.][]{Crocce06}, or simulations with different
dark energy models. In any case, more general modelling of the non-linear effect
will be required for an accurate interpretation of future measurements of the
LSS-CMB cross-correlation.

\section*{ACKNOWLEDGEMENT}
The Millennium simulation used in this paper was carried out by the
Virgo Consortium at the Computing Centre of the Max-Planck Society in
Garching. YC is supported by the Marie Curie Early Stage Training Host
Fellowship ICCIPPP, which is funded by the European Commission. We
thank Raul Angulo for providing the L-BASICC simulation, which was
carried out on the Cosmology Machine at Durham, and for useful
discussions. We also thank Anthony Challinor, Enrique Gaztanaga and
Uros Seljak for comments that allowed us to identify an error in an
earlier version of this paper. This work was supported in part by an
STFC rolling grant. CSF acknowledges a Royal-Society Wolfson Research
Merit Award.

\appendix
\section{Angular Power Spectra}
Here we derive the relationship between the 3-D power spectrum of
gravitational potential fluctuations, $P_{\dot{\Phi}\dot{\Phi}}(k,t)=
(2\pi)^{-3}\langle \dot{\Phi}(k,t) \dot{\Phi}^*(k,t) \rangle $ and the resulting
angular power spectrum of the induced CMB temperature fluctuations.
Expanding the pattern of temperature fluctuations,
$\Delta T(\hat r)/{\bar T_0}$, in terms of spherical harmonics we have
\begin{equation}
a_{lm}=\int  \frac{\Delta T(\hat r)}{\bar T_0} \ Y_{lm}^* (\hat r)\, d\hat r
\end{equation}
which using equation~(\ref{eq1}) becomes
\begin{equation}
a_{lm} =-\frac{2}{c^2}\int Y_{lm}^*  (\hat r) \int_0 ^{t_L}
\dot\Phi(\hat r, t) \, dt \,d\hat r .
\end{equation}
Writing $\dot\Phi(\hat r, t)$ in terms of a Fourier expansion and
using the spherical harmonic expansion of a plane wave,
$
\exp(i\vec k\cdot \vec r)=4\pi\sum_{lm}i^lj_l(kr)Y_{lm}^*(\hat k)Y_{lm}(\hat r)
$
 this becomes
\begin{equation}
\begin{aligned}
a_{lm}&=-\frac{2}{(2\pi)^3c^2}\int Y_{lm}^*  (\hat r) \int_0 ^{t_{L}} \int\dot\Phi(\vec k, t) \exp(i\vec k\cdot \vec r)d\vec k dt d\hat r \\
&=-\frac{2\times 4\pi}{(2\pi)^3c^2}\int Y_{lm}^*  (\hat r) \int_0
 ^{t_{L}} \int\dot\Phi(\vec k, t)\times \\
&\quad \sum_{l'm'}i^lj_l(kr)Y_{l'm'}^*(\hat k)Y_{l'm'}(\hat r)\, d\vec k
\, dt \, d\hat r \\
&=-\frac{1}{\pi^2 c^2} \int_0 ^{t_{L}} \int\dot\Phi(\vec
k, t) i^lj_l(kr)Y_{lm}^*(\hat k) \, d\vec k \, dt .
\end{aligned}
\end{equation}
Hence the angular power spectrum, $C_l$, is given by
\begin{equation}
\begin{aligned}
&C_l \, \delta_{ll^\prime} \delta_{mm^\prime}
\equiv \left\langle a_{lm}a_{l'm'}^* \right\rangle \\
&=\left[\frac{1}{\pi^2 c^2}\right]^2 \Big\langle \int_0
^{r_{L}} \int\dot\Phi(\vec k, r) i^lj_l(kr)Y_{lm}^*(\hat k) \, d\vec k
\, dr \\
& \quad \times \int_0 ^{r_{L}} \int\dot\Phi^*(\vec k', r')
i^{l'}j_l(k'r')Y_{l'm'}(\hat k') d\vec k' dr' \Big\rangle  .
\end{aligned}
\end{equation}
Using  the identity
$ \left\langle \dot\Phi(\vec k)\dot\Phi^*(\vec k') \right\rangle
\equiv (2\pi)^3\delta(\vec k-\vec k')P_{\dot\Phi\dot\Phi}(k)$
and the orthogonality relationship of spherical harmonics
\begin{equation}
\int_{4\pi} Y_{lm}^*(\hat r)Y_{l'm'}(\hat r)\, d\hat r=\delta_{ll^\prime}\delta_{mm^\prime}
\end{equation}
this becomes
\begin{equation}
\begin{aligned}
&C_l \, \delta_{ll^\prime} \delta_{mm^\prime}
=\frac{8}{\pi c^4}\frac{2}{\pi} \int \int_0 ^{r_{L}}\int_0 ^{r_{L}}  P_{\dot\Phi\dot\Phi}(k, r, r') i^lj_l(kr)Y_{lm}^*(\hat k) \\
& \quad \times i^{l'}j_l(kr')Y_{l'm'}(\hat k) \, dr' \, dr \, d\vec k \\
&=\frac{8}{\pi c^4} \int \int_0 ^{r_{L}}\int_0 ^{r_{L}} k^2
P_{\dot\Phi\dot\Phi}(k, r, r')j_l(kr)j_l(kr')dr' dr dk
\delta_{ll^\prime} \delta_{mm^\prime} \\
&C_l=\frac{8}{\pi c^4} \int \int_0 ^{r_{L}}\int_0 ^{r_{L}} k^2
P_{\dot\Phi\dot\Phi}(k, r, r')j_l(kr)j_l(kr')dr' dr dk .
\end{aligned}
\end{equation}
This exact relationship can be simplified by using Limber's
approximation. For small angular separations, $\theta$, at comoving
distance, $r$, the wave number, $k$, can be expressed in terms of its
components parallel and perpendicular to the line of sight and
approximated by $k=\sqrt{k_{\parallel}^2+k_\perp^2}\approx k_\perp$,
where $k_\perp=2\pi/r\theta \approx l/r \gg k_{\parallel}\sim 1/\Delta
r$, namely, the power is dominated by that perpendicular to the
line of sight and there is no correlation between different shells of
$\Delta r$ along the line of sight. Combining this with the
orthogonality relation for spherical Bessel functions,
\begin{equation}
{2 \over \pi} \int k^2 j_l(kr)j_l(kr')dk=\delta(r-r')/r^2, 
\end{equation}
we arrive at
\begin{equation}
C_l \approx \frac{4}{c^4}\int_0 ^{r_{L}}
P_{\dot\Phi\dot\Phi}(k=\frac{l}{r}, r)/r^2 dr.
\end{equation}
\citep[see also][]{Limber54,Kaiser92,Hu00,Verde00}.  
\citet{Verde00} find that the difference between this
approximation and the full calculation is less than $3\%$ at
$l>20$. In this paper we are mainly concerned with even smaller scales
($l>500$) and so we are justified in using Limber's approximation.  We
also tested the difference between using $k=l/r$ and the more accurate
$k=(l+1/2)/r$ \citep{Loverde08} and find very little
difference to our results.

\end{document}